\documentclass[12pt,preprint]{article}
\usepackage{amssymb}
\usepackage{amsmath}
\usepackage{graphics}
\usepackage{epsfig}

\renewcommand{\vec}[1]{{\bf #1}}
\newcommand{\eqb}{\begin{equation}}
\newcommand{\eqe}{\end{equation}}
\newcommand{\dmb}{\begin{displaymath}}
\newcommand{\dme}{\end{displaymath}}

\newcommand{\eab}{\begin{eqnarray}}
\newcommand{\eae}{\end{eqnarray}}

\newcommand{\be}{\begin{equation}}
\newcommand{\ee}{\end{equation}}

\begin{document}

\begin{titlepage}
\begin{flushright} 
\end{flushright}
\vspace{0.6cm}

\begin{center}
\Large{Evolving center-vortex loops}

\vspace{1.5cm}

\large{Julian Moosmann and Ralf Hofmann}
\end{center}
\vspace{1.5cm} 

\begin{center}
{\em Institut f\"ur Theoretische Physik\\ 
Universit\"at Karlsruhe (TH)\\ 
Kaiserstr. 12\\ 
76131 Karlsruhe, Germany}
\end{center}
\vspace{1.5cm}

\begin{abstract}
We consider coarse-graining applied to nonselfintersecting planar
center-vortex loops 
as they emerge in the confining phase of an 
SU(2) Yang-Mills theory. Well-established properties of planar curve-shrinking predict 
that a suitably defined, geometric effective action exhibits (mean-field) critical 
behavior when the conformal limit of circular points is reached. 
This suggests the existence of an asymptotic 
mass gap. We demonstrate that the 
initially sharp mean center-of-mass position in a given ensemble of 
curves develops a variance under the flow as is the case for a 
position eigenstate in free-particle quantum mechanics under unitary
time evolution. A possible application of 
these concepts is an approach to high-$T_c$ superconductivity 
based (a) on the nonlocal nature of the electron 
(1-fold selfintersecting center-vortex loop) and (b) on planar curve-shrinking 
flow representing the decrease in thermal 
noise in a cooling cuprate.       
\end{abstract} 

\end{titlepage}

\section{Introduction}

Four dimensional SU(2) Yang-Mills theory occurs in three phases: a deconfining, a
preconfining, and a confining one. While the former two 
phases possess propagating gauge fields a complete 
decoupling thereof takes place at a Hagedorn 
transition towards the confining phase
\cite{Hofmann2005,GHS2006,Hofmann2007}. 
Namely, by the decay of a preconfining ground state, consisting of 
collapsing magnetic (w.r.t. the gauge fields in the 
defining SU(2) Yang-Mills Lagrangian) flux lines of finite core-size $d$, see also
\cite{NO}, 
into nonselfintersecting or selfintersecting 
center-vortex loops \cite{tHooft1983} 
the mass $m_D$ of the dual gauge 
field diverges. This, in turn, implies $d\to 0$. 
As a consequence, center-vortex loops (CVLs) with nonvanishing selfintersection 
number $N$ become stable solitons in isolation. These solitons are 
classified according to their topology and center charge. That is, for $d\to 0$ the region of negative 
pressure $P$ is confined to the vanishing vortex core, and
the soliton becomes a particle-like ($P=0$) excitation whose stability
is in addition assured topologically by its selfintersection(s).

The purpose of our paper is to investigate the sector with $N=0$ in some
detail. Topologically, there is no reason for the stability of this
sector's excitations, and we will argue that on average and as a consequence 
of a noisy environment a planar CVL with $N=0$ shrinks to 
nothingness within a finite `time'. Here the role of `time' is 
played by a variable measuring the decrease of externally provided
resolving power applied to the system. By `planar CVL' we mean an
embedding of the $N=0$ soliton into a 2D flat and spatial plane. 
For an isolated SU(2) Yang-Mills theory the role of the 
environment is played by the sectors with $N>0$. If the SU(2) theory
under consideration is part of a world subject to additional gauge
symmetries then a portion of such an environment arises from a mixing with
these theories. In any case, a planar CVL at finite length $L$ is
acquiring mass by frequent interactions with the environment after it was 
generated by a process subject to an  
inherent, finite resolution $Q_0$. At the same time, the CVL starts 
shrinking towards a circular point. The latter becomes unresolvable starting
at some finite resolution $Q_*$. That is, all properties that are related to the
existence of extended lines of center flux, observable for
$Q_0\ge Q>Q_*$, do not occur for $Q\le Q_*$, and the CVL vanishes from the spectrum of confining SU(2)
Quantum Yang-Mills theory. Since CVLs with $N>0$ have a finite mass
(positive-integer multiples of the Yang-Mills scale $\Lambda$) we
`observe' a gap in the mass spectrum of the theory when probing the
system with resolution $Q\le Q_*$.      

\begin{figure}
\begin{center}
\leavevmode
\leavevmode
%\epsffile[80 25 534 344]{}
\vspace{4.3cm}
\includegraphics{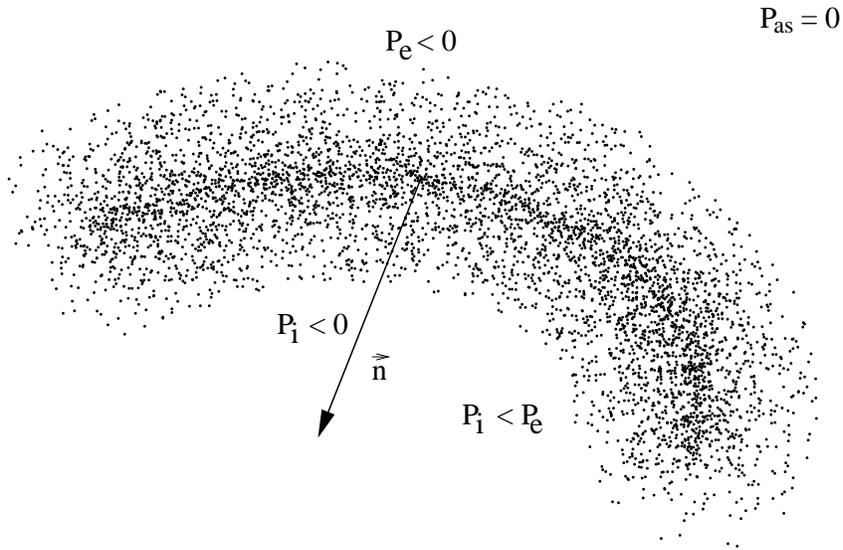}
\end{center}
\caption{\protect{\label{Fig-1}} Highly space-resolved snapshot of an $N=0$ CVL
  curve-sector. The pressure $P_i$ in the region pointed to by the normal
  vector $\vec{n}$ is more negative than the pressure 
$P_e$ thus leading to a motion of the sector along $\vec{n}$.}
\end{figure}

Notice that by embedding an $N=0$ CVL of an isolated SU(2) Yang-Mills theory into a 
flat 2D surface at $m_D<\infty$, $d>0$, a hypothetical observer measuring a positive (negative) 
curvature along a given sector of the vortex line experiences more (less)
negative pressure in the intermediate vicinity of this curve sector 
leading to this sector's motion towards (away from)
the observer, see Fig.\,\ref{Fig-1}. 
%***********************
%************************ 
The speed of this 
motion is a monotonic function of curvature. 
On average, this shrinks the CVL. Alternatively, one may {\sl globally}
consider the limit 
$m_D\to\infty$, $d\to 0$,
that is, the confining phase of an SU(2) Yang-Mills theory, 
but now take into account the effects of an environment which 
{\sl locally} relaxes this limit (by collisions) and thus also induces
curve-shrinking. One possibility to describe this situation is by the 
following flow equation in the (dimensionless) parameter 
$\tau$    
%*********
\eqb
\label{CSt}
\partial_\tau \vec{x}=\frac{1}{\sigma}\,\partial^2_s \vec{x}\,,
\eqe
%**********
where $s$ is arc length, $\vec{x}$ is a point on the planar CVL, and 
$\sigma$ is a string tension effectively 
expressing the distortions induced by the environment. After a rescaling,
$\hat{x}\equiv\sqrt{\sigma}\vec{x}$, $\xi=\sqrt{\sigma}s$,
Eq.\,(\ref{CSt}) assumes the following form
%*********
\eqb
\label{CSta}
\partial_\tau \hat{x}(u,\tau)=\partial^2_\xi \hat{x}=k(u,\tau)\vec{n}(u,\tau)\,,
\eqe
%**********
where $u$ is a (dimensionless) curve parameter ($d\xi=|\partial_u\hat{x}|\,du$), $\vec{n}$ 
the (inward-pointing) Euclidean unit normal, $k$ the
scalar curvature, defined as 
%*********
\eqb
\label{curvdef}
k\equiv\left|\partial^2_\xi \hat{x}\right|
=\left|\frac{1}{|\partial_u\hat{x}|}\partial_u\left(\frac{1}{|\partial_u\hat{x}|}\partial_u\hat{x}\right)\right|\,,
\eqe
%**********
$|\vec{v}|\equiv\sqrt{\vec{v}\cdot\vec{v}}$, and $\vec{v}\cdot\vec{w}$ 
denotes the Euclidean scalar product of the vectors $\vec{v}$ and
$\vec{w}$ lying in the plane. In the following we resort to a slight abuse of notation by
using the same symbol $\hat{x}$ for the functional dependence on $u$ or
$\xi$. 

It is worth mentioning that Eq.\,(\ref{CSta}) expresses a special case
of the local condition that the rate of decrease of the (dimensionless) 
curve length $L(\tau)=\int_0^{L(\tau)} d\xi=\int_0^{2\pi}
du\,\left|\partial_u\hat{x}(u,\tau)\right|$ is 
maximal w.r.t. a variation of the direction of the velocity
$\partial_\tau\hat{x}$ of a given point on the curve at fixed 
$|\partial_\tau\hat{x}|$ \cite{Smith}:
%**********
\eqb
\label{roc}
\frac{dL(\tau)}{d\tau}=-\int_0^{L(\tau)} d\xi\,k\vec{n}\cdot\partial_\tau
\hat{x}\,.
\eqe
%*********
The 1D heat-equation (\ref{CSta}) 
is well understood mathematically \cite{GageHamilton,Grayson} and
represents the 1D analogue of the Ricci equation 
describing curvature-ho\-moge\-ni\-zation of 3-manifolds 
\cite{PerelmanI,PerelmanII,PerelmanIII} thus proving the geometrization
conjecture \cite{Thurston}. 

The present work interpretes the 
shrinking of closed and 2D-flat embedded (planar) curves as a Wilsonian 
renormalization-group evolution governed by an effective 
action which is defined purely in geometric terms. In the presence of an
environment represented by the parameter $\sigma$, 
this action possesses a natural decomposition into a conformal and a
nonconformal factor. One of our goals is to show 
that the transition to the conformal limit of vanishing 
mean curve length really is a critical phenomenon 
characterized by a mean-field exponent if a suitable parameterization of
the effective action is used. To see this, various 
initial conditions are chosen to generate an 
ensemble whose partition function is 
invariant under curve-shrinking. A (second-order) 
phase transition is characterized by the critical behavior of the 
coefficient associated with the nonconformal factor in the effective action. That is, in the presence 
of an environment the (nearly massless) $N=0$ sector in confining SU(2)
Yang-Mills dynamics, 
generated during the Hagedorn transition, practically disappears after a finite 
`time' leading to an asymptotic mass gap. We also believe that $N=0$ 
CVLs play the role of Majorana neutrinos in physics: 
Their disappearance after a finite time and the absence of antiparticles
would be manifestations of lepton-number violation \cite{KKG} forbidden in the present
Standard Model of Particle Physics and may provide for an explanation to the
solar neutrino problem alternative to the oscillation scenario, see also \cite{BosiCavalleri2002}.    

In Sec.\,\ref{MP} we provide information on proven properties 
of curve-shrinking evolution. The philosophy 
underlying the statistics of geometric fluctuations in the $N=0$ 
sector is elucidated in Sec.\,\ref{PF}. In Sec.\,\ref{RGF} we present
our results for the renormalization-group flow of the effective 
action, give an interpretation, and compute the evolution of a local
quantity. Finally, in Sec.\,\ref{S} 
we summarize our work and give an outlook to the 
$N=1$ case which we suspect is relevant for high-$T_c$
superconductivity.                 

\section{Prerequisites on mathematical results\label{MP}}

In this section we provide knowledge on the properties of the shrinking
of embedded (nonselfintersecting) curves 
in a plane \cite{GageHamilton,Grayson}. It is important to stress that 
only for curve-shrinking in a plane are the following results
valid. To restrict the motion of a 
CVL to a plane is a major assumption, and additional physical 
arguments must be provided for its validity. 

The properties of the $\tau$-evolution of smooth, embedded, and closed
curves $\hat{x}(u,\tau)$ subject to Eq.\,(\ref{CSta}) 
were investigated in \cite{GageHamilton} for the purely convex case 
and in \cite{Grayson} for the general case. 
The main result of \cite{Grayson} is 
that an embedded curve with finitely many points of 
inflection remains embedded and smooth when 
evolving under Eq.\,(\ref{CSta}) and that such a curve flows to a circular 
point for $\tau\nearrow T$ where $0<T<\infty$. That is, asymptotically the curve converges (w.r.t. the
$C^\infty$-norm) to a shrinking circle: $\lim_{\tau\to T} L(\tau)=0$ and
$\lim_{\tau\to T} A(\tau)=0$, $A$ being the (dimensionless) area
enclosed by the curve, 
such that the isoperimetric ratio $\frac{L^2(\tau)}{A(\tau)}$ approaches the value $4\pi$
from above. 
%***********************
\begin{figure}
\begin{center}
\leavevmode
\leavevmode
%\epsffile[80 25 534 344]{}
\vspace{5.3cm}
\includegraphics{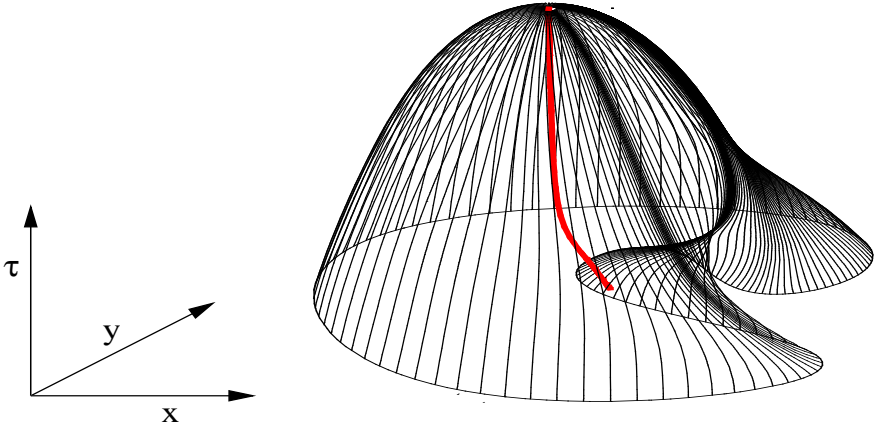}
\end{center}
\caption{\protect{\label{Fig-2}} Plot of the 
evolution of a planar CVL under Eq.\,(\ref{CSta}). The thick
central line depicts the graph of the CVL's `center of mass'. The
flow is started at $\tau=0$ and ends at $\tau=T$.}
\end{figure}
%************************ 
For later use, we present the following two identities, see Lemma 3.1.2
and Lemma 3.1.7 in \cite{GageHamilton}:
%***********
\eqb
\label{Levol}
\partial_\tau L=-\int_0^L d\xi\,k^2=-\int_0^{2\pi}du\,|\partial_u\hat{x}|k^2\,,
\eqe
%*********** 
%***********
\eqb
\label{Aevol}
\partial_\tau A=-2\pi\,.
\eqe
%*********** 
Setting $A(\tau=0)\equiv A_0$, the solution to Eq.\,(\ref{Aevol}) is
%********
\eqb
\label{solAevol}
\frac{A(\tau)}{A_0}=1-\frac{2\pi\tau}{A_0}\,.
\eqe
%********
By virtue of Eq.\,(\ref{solAevol}) the critical value $T$ is related 
to $A_0$ as
%*******
\eqb
\label{TandA0}
T=\frac{A_0}{2\pi}\,.
\eqe
%*******

\section{Geometric partition function \label{PF} }

We now wish to interprete curve-shrinking as a Wilsonian
renormalization-group flow taking place in the $N=0$ 
planar CVL sector. A partition function, 
defined as a statistical average (according to a suitably defined weight) over $N=0$ 
CVLs, is to be left invariant under a decrease of the 
resolution determined by the flow parameter $\tau$. 
Notice that, physically, $\tau$ is interpreted as a strictly
monotonic decreasing (dimensionless) 
function of a ratio $\frac{Q}{Q_0}$ where $Q$ ($Q_0$) are mass 
scales associated with an actual (initial) resolution applied to the
system. 

To device a geometric ansatz for the 
effective action $S=S[\hat{x}(\tau)]$, which is a functional of the
curve $\hat{x}$ representable in terms 
of integrals over local densities in $\xi$ (reparametrization
invariance), 
the following reflection on symmetries is in order.  
(i) scaling symmetry $\hat{x}\to \lambda\hat{x}\,,\ \ \lambda\in{\mathbf
  R}_+$: For both, $\lambda\to\infty$ and $\lambda\to 0$, implying
$\lambda L\to\infty$ and $\lambda L\to 0$ at fixed $L$, the action $S$ should 
be invariant under further finite rescalings (decoupling of the fixed
length scale $\sigma^{-1/2}$), 
(ii) Euclidean point symmetry of the plane (rotations, translations
and reflections about a given axis): Sufficient but not necessary for
this is a 
representation of $S$ in terms of integrals over 
scalar densities w.r.t. these 
symmetries. That is, the action density should be expressible as a
series involving products of Euclidean scalar products of $\frac{\partial^n}{\partial
    \xi^n}\hat{x}\,,\ \ n\in\mathbf{N}_+\,,$ or constancy. 
However, an exceptional scalar integral over a nonscalar density can be 
deviced. Consider the area $A$, calculated as 
%*********
\eqb
\label{area}
A=\left|\frac12\,\int_0^{2\pi} d\xi\,\hat{x}\cdot\vec{n}\right|\,.
\eqe
%*********  
The density $\hat{x}\cdot\vec{n}$ in Eq.\,(\ref{area}) is not a scalar under
translations. 

We now resort to a 
factorization ansatz as 
%*********
\eqb
\label{effectactdef}
S=F_c\times F_{nc}\,,
\eqe
%**********
where in addition to
Euclidean point symmetry $F_c$ ($F_{nc}$) is (is not) invariant 
under $\hat{x}\to \lambda\hat{x}$. In principle, 
infinitely many operators can be defined 
to contribute to $F_c$. Since the 
evolution generates circles for $\tau\nearrow T$ higher derivatives of 
$k$ w.r.t. $\xi$ rapidly converge to zero \cite{GageHamilton}. We expect this to be true 
also for Euclidean scalar products involving higher 
derivatives $\frac{\partial^n}{\partial \xi^n}\hat{x}$. 
To yield conformally invariant expressions such integrals need to be 
multiplied by powers of $\sqrt{A}$ and/or $L$ or the inverse of 
integrals involving lower derivatives. At this stage, we are 
not capable of constraining the expansion in derivatives by additional physical or
mathematical arguments. To be pragmatic, 
we simply set $F_c$ equal to the isoperimetric ratio:
%*********
\eqb
\label{explFc}
F_c(\tau)\equiv\frac{L(\tau)^2}{A(\tau)}\,.
\eqe
%*********
We conceive the nonconformal factor $F_{nc}$ in $S$ as a formal expansion in inverse powers of 
$L$. Since we regard the renormalization-group evolution of the
effective action as induced by the flow of an ensemble of curves, where the evolution of each member is
dictated by Eq.\,(\ref{CSta}), we allow for an explicit $\tau$
dependence of the coefficient $c$ of the lowest nontrivial power
$\frac{1}{L}$. In principle, this sums up the contribution to $F_{nc}$ of certain 
higher-power operators which do not exhibit an explicit $\tau$
dependence. Hence we make the following ansatz
%********
\eqb
\label{explFnc}
F_{nc}(\tau)=1+\frac{c(\tau)}{L(\tau)}\,.
\eqe
%*******
The initial value $c(\tau=0)$ is determined from a physical 
boundary condition such as the mean length $\bar{L}$ at $\tau=0$ which determines
the mean mass $\bar{m}$ of a CVL as $\bar{m}=\sigma\bar{L}$. 

For latter use we investigate the behavior of $F_{nc}(\tau)$ for
$\tau\nearrow T$ for an ensemble consisting of a single curve only and 
require the independence of the `partition function' under changes in
$\tau$. Integrating Eq.\,(\ref{Levol}) in the vicinity of 
$\tau=T$ under the boundary condition that $L(\tau=T)=0$, 
we have
%************
\eqb
\label{soltau=T}
L(\tau)=\sqrt{8}\pi\,\sqrt{T-\tau}\,.
\eqe
%************
Since $F_c(\tau\nearrow T)=4\pi$ independence of the `partition
function' under the flow in $\tau$ implies that 
%*******
\eqb
\label{ctau=T}
c(\tau)\propto\sqrt{T-\tau}\,.
\eqe
%*******
That is, $F_{nc}$ approaches constancy for $\tau\nearrow T$ which brings
us back to the conformal limit by a finite renormalization of the
conformal part $\int F_c$ of the action. In this parameterization of $S$, $c(\tau)$ can thus be regarded
as an order parameter for conformal symmetry with mean-field 
critical exponent.              

\section{Renormalization-group flow \label{RGF} } 

\subsection{Effective action\label{EA}}

\begin{figure}
\begin{center}
\leavevmode
\leavevmode
%\epsffile[80 25 534 344]{}
\vspace{4.3cm}
\includegraphics{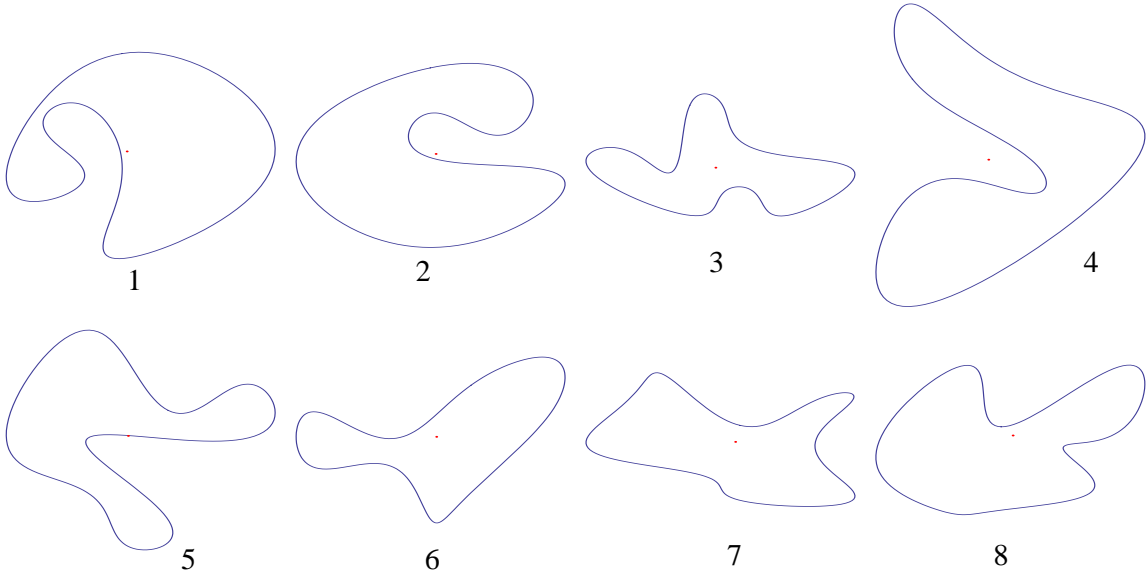}
\end{center}
\caption{\protect{\label{Fig-3}}Initial curves contributing to the
  ensembles $E_M$, see text. Points locate the respective positions of
the centers of mass.}
\end{figure}

Let us now numerically investigate the effective action
$S[\hat{x}(\tau)]$ resulting from a 
partition function $Z$ w.r.t. a nontrivial ensemble $E$. The latter is defined as the average
%*******
\eqb
\label{PartZM}
Z=\sum_{i} \exp\left(-S[\hat{x}_i(\tau)]\right)
\eqe
%*******
over the ensemble $E=\{\hat{x}_1,\cdots\}$. 
Let us denote by $E_M$ an ensemble consisting of $M$ curves where 
$E_M$ is obtained from $E_{M-1}$ by adding a new curve $\hat{x}_M(u,\tau)$. In
Fig.\,\ref{Fig-3} eight initial curves are depicted which in this way generate the
ensembles $E_M$ for $M=1,\cdots,8$. 
%***********************
%************************  

We are interested 
in a situation where all curves in $E_M$ shrink to a point at the same
value $\tau=T$. Because of Eqs.\,(\ref{solAevol}) 
and (\ref{TandA0}) we thus demand that at $\tau=0$ all curves in 
$E_M$ initially have the same area $A_0$. The effective action $S$ in
Eq.\,(\ref{effectactdef}) (when associated with the ensemble $E_M$ we
will denote it as $S_M$) is determined by the function $c_M(\tau)$,
compare with Eq.\,(\ref{explFnc}), whose 
flow follows from the requirement of $\tau$-independence of $Z_M$:
%********
\eqb
\label{renflow}
\frac{d}{d\tau}Z_M=0\,.
\eqe 
%*******
This is an implicit, first-order ordinary differential equation 
for $c(\tau)$ which needs to be supplemented with an initial condition
$c_{0,M}=c_M(\tau=0)$. A natural initial condition is to 
demand that the quantity 
%********
\eqb
\label{barL}
\bar{L}_M(\tau=0)\equiv\frac{1}{Z_M(\tau=0)}\sum_{i=1}^M
L[\hat{x}_i(\tau=0)]\,\exp\left(-S_M[\hat{x}_i(\tau=0)]\right)
\eqe
%*******
coincides with the geometric mean $\tilde{L}_M(\tau=0)$ defined as 
%**********
\eqb
\label{geommean}
\tilde{L}_M(\tau=0)\equiv\frac{1}{M}\sum_{i=1}^M L[\hat{x}_i(\tau=0)]\,.
\eqe
%********** 
From $\bar{L}_M(\tau=0)=\tilde{L}_M(\tau=0)$ a value for $c_{0,M}$
follows. We also have considered a modified factor 
$F_{nc}(\tau)=1+\frac{c(\tau)}{A(\tau)}$ in
Eq.\,(\ref{effectactdef}). In this case the choice of initial condition 
$\bar{L}_M(\tau=0)=\tilde{L}_M(\tau=0)$ leads to $F_{nc}(\tau)\equiv
0$. While the geometric effective action thus is profoundly different 
for such a modification of $F_{nc}(\tau)$ physical results such 
as the evolution of the variance of center-of-mass position agree
remarkably well, see Sec.\,\ref{COM}. That is, 
the geometric effective action itself is not a physical 
object. Rather, going from one ansatz for $S_M$ to another describes 
a particular way of redistributing the weight in the
ensemble which seems to have no significant impact on the physics. 
This is in contrast to quantum field theory and conventional
statistical mechanics where the action in principle is related to the 
physical properties of a given member 
of the ensemble. 

\begin{figure}
\begin{center}
\leavevmode
\leavevmode
%\epsffile[80 25 534 344]{}
\vspace{7.3cm}
\includegraphics{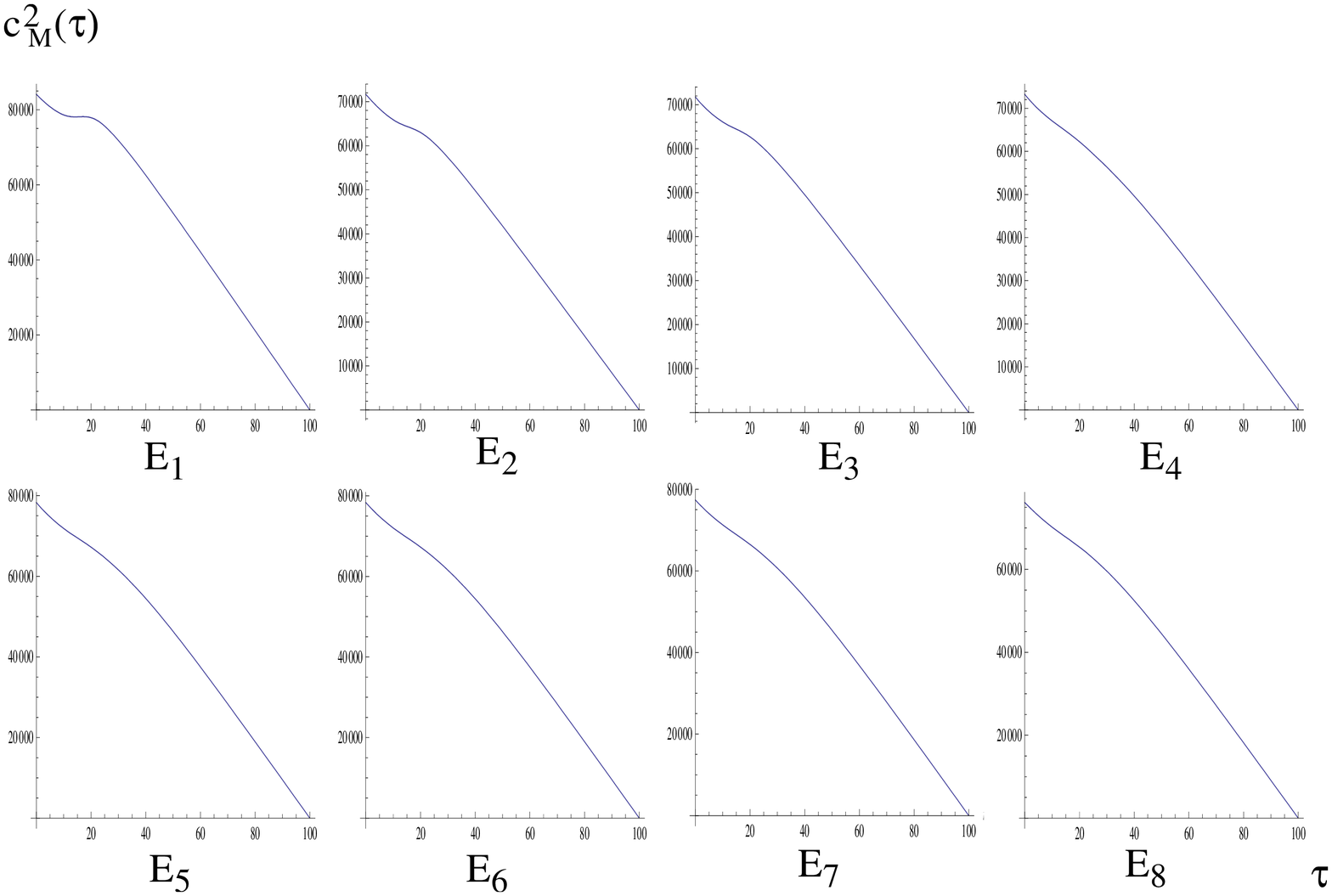}
\end{center}
\caption{\protect{\label{Fig-4}}The square of the coefficient $c_M(\tau)$
  entering the effective action of Eq.\,(\ref{effectactdef}) by virtue
  of Eq.\,(\ref{explFnc}) for various ensemble sizes. 
Notice the early onset of the linear drop of $c^2_M(\tau)$ and the 
saturation in $M$ for $M\ge 5$. The slope of $c^2_M(\tau)$ near 
$\tau=T$ does not depend on $c_{0,M}^2\equiv c^2_M(\tau=0)$ and thus not on the initial
choice of $\bar{L}$. }
\end{figure}

For the curves depicted in Fig.\,\ref{Fig-3} 
we make the convention that $A_0\equiv 2\pi\times 100$. It then follows that $T=100$ by
Eq.\,(\ref{TandA0}). The dependence $c_M^2(\tau)$ 
is plotted in Fig.\,\ref{Fig-4}. According
to Fig.\,\ref{Fig-4} it seems
that the larger the ensemble the closer $c_M^2(\tau)$ 
to the evolution of a single 
circle of initial radius $R=\sqrt{\frac{A_0}{\pi}}$. That is, for
growing $M$ the function $c_M^2(\tau)$ approaches the form 
%*****
\eqb
\label{cas}
c^2_{\tiny\mbox{as},M}(\tau)=k_M(T-\tau)\,,
\eqe
%******
where the slope $k_M$ depends on the 
strength of deviation from circles of 
the representatives in the ensemble $E_M$ at $\tau=0$, that is, on the
variance $\Delta L_M$ at a given value $A_0$.  
Physically speaking, the value $\tau=0$ is associated 
with a certain initial resolution of the measuring 
device (the strictly monotonic function $\tau(Q)$, $Q$ being a physical scale such as energy
or momentum transfer, expresses the characteristics 
of the measuring device and the measuring process), the value of $A_0$ describes the 
strength of noise associated with the environment ($A_0$ 
determines how fast the conformal limit of circular points is reached), and the values 
of $c_{0,M}$ and $k_M$, see Eq.\,(\ref{cas}), are associated 
with the conditions at which the to-be-coarse-grained 
system is prepared. Notice that this interpretation 
is valid for the 
action $S_M=\frac{L(\tau)^2}{A(\tau)}\left(1+\frac{c_M(\tau)}{L(\tau)}\right)$
only.          
%***********************
%************************     
   
\subsection{Statistical uncertainty of center of mass position\label{COM}}

We are now in a position to 
compute the flow of a more local `observable', namely, the mean
`center-of-mass' (COM) position in a given ensemble and the statistical 
variance of the COM position. The COM position
$\hat{x}_{\tiny\mbox{COM}}$ of a given curve $\hat{x}(\xi,\tau)$ is defined as: 
%***********
\eqb
\label{COMdef}
\hat{x}_{\tiny\mbox{COM}}(\tau)=\frac{1}{L(\tau)}\int_0^{L(\tau)}
d\xi\,\hat{x}(\xi,\tau)\,.
\eqe
%*********** 
We will below present only results on the statistical 
variance of the COM position. 

Let as assume that at $\tau=0$ the ensembles $E_M$ are 
modified such that a translation is applied to 
each representative letting its COM position coincide with the 
origin. Recall that such a modification $E_M\to E^\prime_M$ does not 
alter the (effective) action (Euclidean point symmetry). That is, at $\tau=0$ 
the statistical variance in the position of the COM is prepared to be 
nil, physically corresponding to an infinite resolution applied to the
system by the measuring device. 

\begin{figure}
\begin{center}
\leavevmode
\leavevmode
%\epsffile[80 25 534 344]{}
\vspace{7cm}
\includegraphics{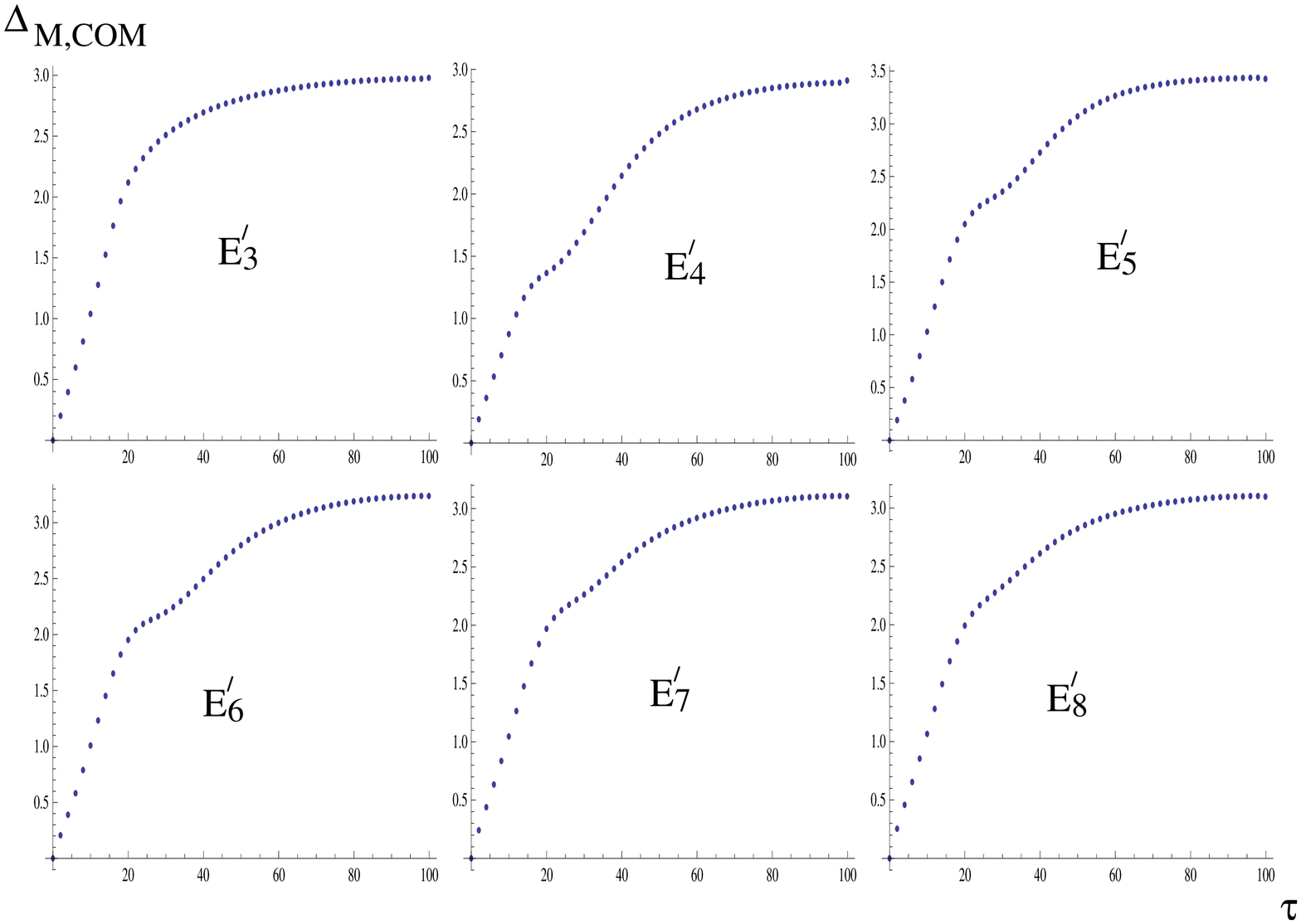}
\end{center}
\caption{\protect{\label{Fig-5}} Plots of 
$\Delta_{M,\tiny\mbox{COM}}(\tau)$ for $M=3,\cdots,8$ when evaluated
with the action
$S_M=\frac{L(\tau)^2}{A(\tau)}\left(1+\frac{c_M(\tau)}{L(\tau)}\right)$. Notice the rapid
generation of an uncertainty in the COM position under the flow 
and its saturation when approaching the conformal limit $\tau\nearrow
T$. There also is a saturation of this limiting value with a growing 
ensemble size.}
\end{figure}

\begin{figure}
\begin{center}
\leavevmode
\leavevmode
%\epsffile[80 25 534 344]{}
\vspace{8.4cm}
\includegraphics{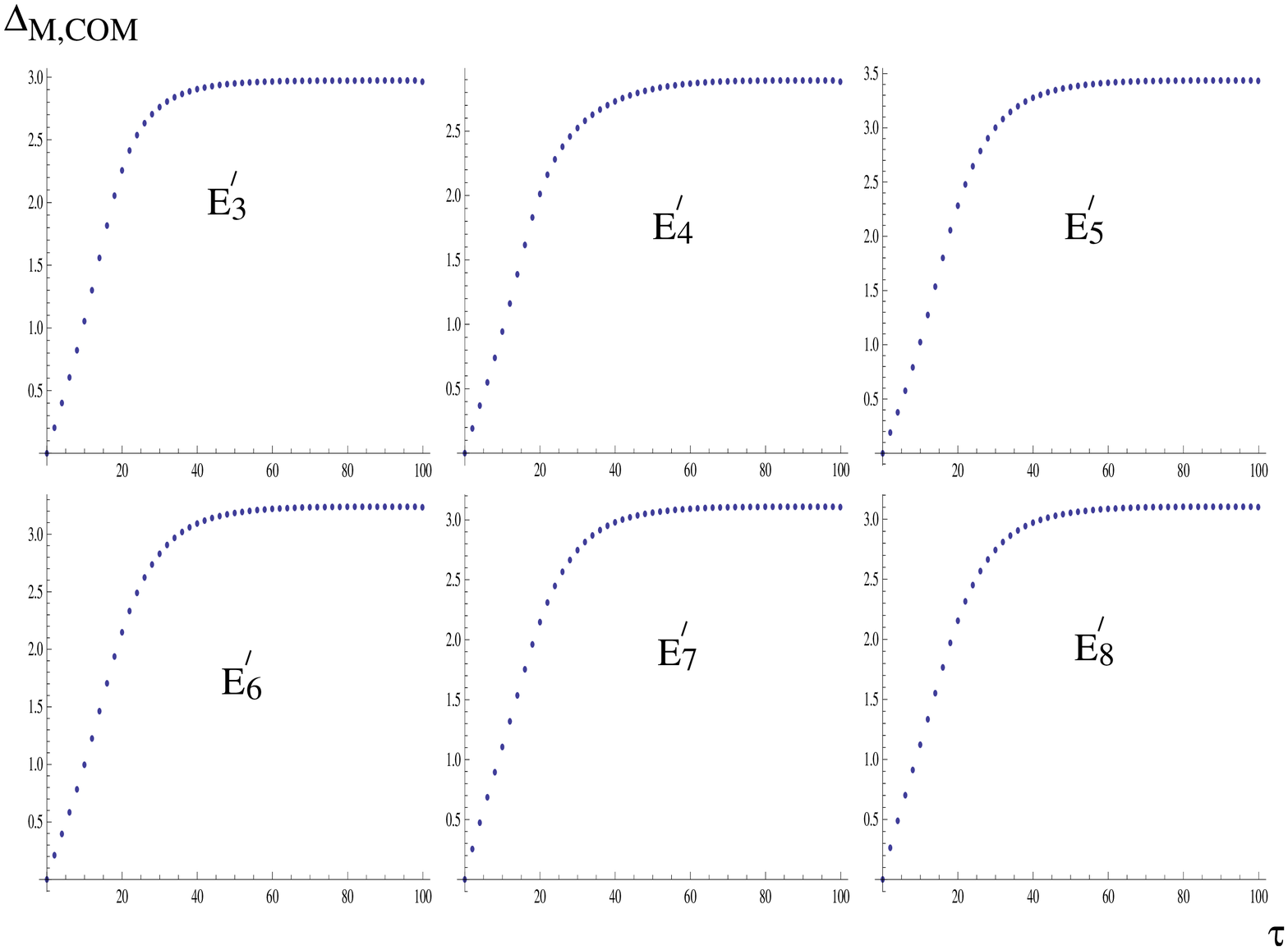}
\end{center}
\caption{\protect{\label{Fig-5b}} Plots of 
$\Delta_{M,\tiny\mbox{COM}}(\tau)$ for $M=3,\cdots,8$ when evaluated
with the action
$S_M=\frac{L(\tau)^2}{A(\tau)}\left(1+\frac{c_M(\tau)}{A(\tau)}\right)$. 
Notice the qualitative agreement with the results displayed in
Fig.\,\ref{Fig-5}.}
\end{figure}

The mean COM position $\bar{\hat{x}}_{\tiny\mbox{COM}}$ over ensemble
$E^\prime_M$ is defined as
%*******
\eqb
\label{menacom}
\bar{\hat{x}}_{\tiny\mbox{COM}}(\tau)\equiv\frac{1}{Z_M}\sum_{i=1}^M
\hat{x}_{\tiny\mbox{COM},i}(\tau)\exp\left(-S_M[\hat{x}_i(\tau)]\right)\,.
\eqe
%*******
The scalar statistical deviation 
$\Delta_{M,\tiny\mbox{COM}}$ of $\bar{\hat{x}}_{\tiny\mbox{COM}}$ over the
ensemble $E^\prime_M$ is defined as
%*******
\eqb
\label{statvar}
\Delta_{M,\tiny\mbox{COM}}(\tau)\equiv \sqrt{\mbox{var}_{M,\tiny\mbox{COM};x}(\tau)+
\mbox{var}_{M,\tiny\mbox{COM};y}(\tau)}\,,
\eqe
%*******
where 
%*******
\eab
\label{vardef}
\mbox{var}_{M,\tiny\mbox{COM};x}&\equiv&\frac{1}{Z_M}\sum_{i=1}^M
\left(x_{\tiny\mbox{COM},i}(\tau)
-\bar{x}_{\tiny\mbox{COM}}(\tau)\right)^2\,\exp\left(-S_M[\hat{x}_i(\tau)]\right)\nonumber\\ 
&=&-\bar{x}^2_{\tiny\mbox{COM}}(\tau)+\frac{1}{Z_M}\sum_{i=1}^M x^2_{\tiny\mbox{COM},i}(\tau)\,
\exp\left(-S_M[\hat{x}_i(\tau)]\right) 
\eae
%**********
and similarly for the coordinate $y$. In Fig.\,\ref{Fig-5} plots of
$\Delta_{M,\tiny\mbox{COM}}(\tau)$ are shown when
$\Delta_{M,\tiny\mbox{COM}}(\tau)$ is evaluated over the ensembles
$E^\prime_3,\cdots,E^\prime_8$ with the action 
%********
\dmb
S_M=\frac{L(\tau)^2}{A(\tau)}\left(1+\frac{c_M(\tau)}{L(\tau)}\right)
\dme
%********
and subject to the initial condition
$\bar{L}_M(\tau=0)=\tilde{L}_M(\tau=0)$. In Fig.\,\ref{Fig-5b} the
according plots of $\Delta_{M,\tiny\mbox{COM}}(\tau)$ are depicted as 
obtained with the action
%********
\dmb 
S_M=\frac{L(\tau)^2}{A(\tau)}\left(1+\frac{c_M(\tau)}{A(\tau)}\right)
\dme
%********
and subject to the initial condition
$\bar{L}_M(\tau=0)=\tilde{L}_M(\tau=0)$. In this case, one has
$c_M(\tau)=-A(\tau)$ leading to equal weights for each curve in
$E^\prime_M$. 
%***********************
%************************  
%***********************
%************************ 

\subsection{Quantum mechanical versus  statistical uncertainty}  

In view of the results obtained in Sec.\,\ref{COM} we would say that an 
ensemble of evolving planar CVLs in the $N=0$ sector qualitatively resembles the
quantum mechanics of a free point particle\footnote{It
is no relevance at this point whether this particle carries spin or
not.} of mass $m$ in 1D. Namely, an initially localized square
of the wave
function $\psi$ with 
$|\psi(\tau=0,x)|^2\propto \exp\left[-\frac{x^2}{a_0^2}\right]$, where
$\Delta x(\tau=0)=a_0$, according 
to unitary time evolution in quantum mechanics evolves as
$|\psi(\tau,x)|^2=|\exp\left[-i\frac{H\tau}{\hbar}\right]\psi(\tau=0,x)|^2\propto
\exp\left[-\frac{(x-\frac{p}{m}\tau)^2}{a^2(\tau)}\right]$ 
where $H=\frac{p^2}{2m}$ is the free-particle Hamiltonian, $p$ is the spatial momentum, and $a(\tau)\equiv
a_0\sqrt{1+\left(\frac{\hbar\tau}{m a_0^2}\right)^2}$. In agreement with
Heisenberg's uncertainty relation one has during the process that $\Delta x\Delta
p=\frac{\hbar}{2}\sqrt{1+\left(\frac{\tau\hbar}{m
      a_0^2}\right)^2}\ge\frac{\hbar}{2}$. Time evolution 
in quantum mechanics and the process of coarse-graining in a statistical
system describing planar CVLs share the property that in 
both systems the 
evolution generates out
of a small initial position uncertainty (corresponding to a large
initial resolution $\Delta p$) a larger 
position uncertainty in the course of the evolution. Possibly, future
development will show that interference effects in quantum mechanics can be traced back 
to the nonlocal nature of the degrees of freedom (CVLs) entering a
statistical 
partition function.

\section{Summary, conclusions, and outlook\label{S}}

\subsection{Summary of present work}

In this exploratory article an attempt has been undertaken 
to interprete the effects of an environment on 2D 
planar center-vortex loops, as they emerge in the confining phase of an
SU(2) Yang-Mills theory, in terms of a Wilsonian renormalization-group 
flow carried by purely geometric entities. Our (mainly numerical) 
analysis uses established mathematics on the shrinking of embedded 
curves in the plane. In the case of nonintersecting 
CVLs ($N=0$) the role of the environment is played by the 
entirety of all sectors with $N>0$ and possibly an explicit 
environment. In a particular parametrization of the effective action 
we observe critical behavior as the limit of circular points is approached during the
evolution. That is, planar $N=0$ CVLs on average disappear from the
spectrum for resolving powers smaller than a critical, finite
value. Using this formalism to compute the evolution of the mean values of local 
observables, such as the center-of-mass position, a 
behavior is generated that qualitatively resembles the associated
unitary evolution in quantum mechanics. We also have found evidence 
that this situation is practically not altered when changing the ansatz for the 
effective action.    

\subsection{Outlook on strongly correlated electrons in a plane}

\begin{figure}
\begin{center}
\leavevmode
\leavevmode
%\epsffile[80 25 534 344]{}
\vspace{3.4cm}
\includegraphics{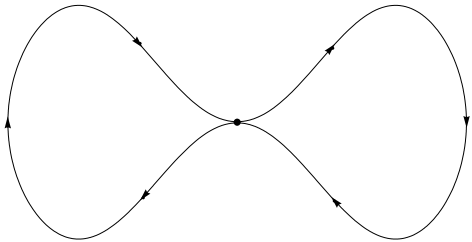}
\end{center}
\caption{\protect{\label{Fig-6}} An isolated CVL with $N=1$, which, when
  associated with the confining phase of an SU(2) Yang-Mills theory of
  scale $\Lambda\sim m_e=511\,$keV, is interpreted as an electron or 
positron. The arrows depict one out of two possible directions of center flux responsible for
the magnetic moment (spin), the intersection point is the location of mass and electric charge.}
\end{figure}

Let us conclude this article with a somewhat 
speculative outlook on planar $N=1$ 
CVLs. Setting the Yang-Mills scale (mass of the intersection in
the CVL) of the associated 
SU(2) theory equal to the electron mass, this 
soliton is interpreted as an electron or positron, 
see \cite{Hofmann2005,Zpinch,GiacosaHofmann2005} for a more
detailed discussion on the viability of such an assignment and
Fig.\,\ref{Fig-6} for a display. Important for our 
purpose are the facts that the two-fold  directional degeneracy of the
center flux represents the two-fold degeneracy of the spin 
projection and that a large class of curve deformations (shifts 
of the intersection point) leaves the
mass of the soliton invariant in the free case. 
%***********************
%************************ 

It is a remarkable fact that a high level of mathematical understanding 
exists for the behavior of curve-shrinking in a 2D plane
\cite{GageHamilton,Grayson} even in the case of one selfintersection \cite{GraysonII}. Incidentally, 2D quantum systems exist in 
nature which exhibit unconventional behavior. 
Specifically, the phenomenon of high-$T_c$ superconductivity 
appears to be strongly related to the two-dimensionality 
of electron dynamics as it is enabled by rare-earth doping 
of cuprate materials \cite{Muller1986}. Apparently, 
the Coulomb repulsion between the electrons moving in the would-be valence
band within the cuprate planes of high-$T_c$ superconductors (Mott
insulators) is effectively screened by the interplane environment also providing for the very 
existence of these electrons by doping. The question then 
is how the long-range order of electronic 
spins, which at given (optimal) doping and at a 
sufficiently low temperature leads to superconductivity, 
emerges within the cuprate planes. As it seems, quantum Monte Carlo 
simulations of a transformed Hubbard model ($t-J$ model) subject to
Gutzwiller projection yield quantitative explanations of 
a number of experimental results related to the existence
of the pseudogap phase (Nernst effect, nonlinear diamagnetic 
susceptibility), see \cite{Anderson2005} and references therein.

We would here like to offer a sketch of an 
alternative approach to high-$T_c$ superconductivity being well 
aware of our ignorance on the details 
of present-day research in this field. The key idea 
is already encoded in Fig.\,\ref{Fig-6}. 
Namely, according to confining
SU(2) Yang-Mills theory the electron is a 
nonlocal object with the physics of its charge localization being
only loosely related to the physics of its magnetic moment (spin): The
magnetic moment, carried by the core of the flux line, microscopically manifested
by (oppositely) moving (opposite) electric charges, receives
contributions from vortex sectors that are spatially far separated (on
the scale of the diameter of the intersection point) from the location
of the isolated electric charge. This suggests that 
in certain physical circumstances, where the ordering effect 
of interacting vortex lines becomes important, the postulate of a spinning
point particle fails to describe reality.
%***********************
\begin{figure}
\begin{center}
\leavevmode
\leavevmode
%\epsffile[80 25 534 344]{}
\vspace{5.3cm}
\includegraphics{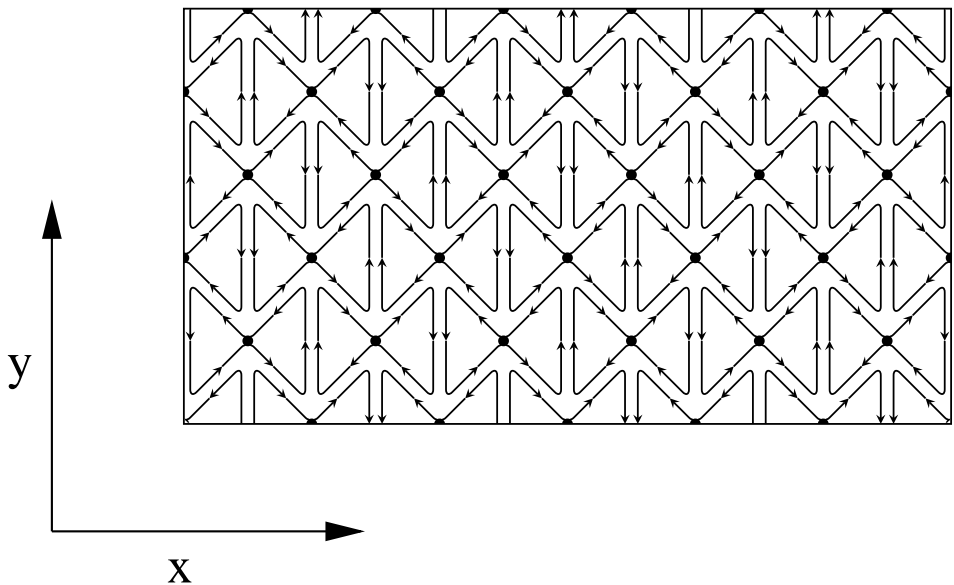}
\end{center}
\caption{\protect{\label{Fig-7}} An array of strongly correlated
  electrons in the plane possibly representing the superconducting state
  in a cuprate. The equally directed center flux in 
adjacent vortex sections provides for an attractive force 
(Ampere's law) at intermediate distances. For a given electron, out of six neighboring 
vortex sectors there are four sectors with attraction and two sectors with repulsion.  
At short distance there is repulsion since an overlap of 
CVL sections, leading to new intersection points each of mass 
$\sim m_e$, is topologically forbidden. Thus there is a typical 
equilibrium configuration contributing to long-range order in the 2D system. If the 
externally provided resolution (temperature) falls below a 
critical value then statistical fluctuations of the position 
of an intersection point relative to another one (the location of the electronic
charge) will vanish. That is, electrons no longer can disperse 
energy provided by the heat bath (phonons, spin fluctuations) and 
thus provide for a 2D material free of electric resistivity.}
\end{figure}
%************************ 

Concerning the strong correlations in 2D electron dynamics responsible
for high-$T_c$ superconductivity we imagine a situation as
depicted in Fig.\,\ref{Fig-7}. Each electron's 
spin in the plane interacts with the spin of its neighbors as 
follows. Equally directed electric fluxes (dually interpreted 
center fluxes of SU(2)$_e$) attract one another, and 
there is attraction for four out of six vortex sectors defined by 
the neighboring electrons (Ampere's law) while two vortex sectors
experience repulsion\footnote{That the electric 
charges of confining SU(2)$_e$ are seen by the photon 
is a consequence of the mixing between the corresponding two 
gauge groups: SU(2)$_e$ and SU(2)$_{\tiny\mbox{CMB}}$
\cite{Hofmann2005,GiacosaHofmann2005}.}. 
Notice that for a given electron two of the adjacent electrons exhibit equal spin projection while
four of the adjacent electrons have opposite spins. This is in agreement
with the observation that high-$T_c$ superconductivity is an effect 
not related to $s$-wave Cooper-pair condensation. 

An overlap of vortex sectors, hypothetically leading 
to the creation of extra intersection points, is topologically forbidden. 
That is, the fluctuations in the energy density of the system are far 
to weak to create an intersection point of mass 
$m_e=511\,$keV. Therefore, at very small spatial separation 
repulsion must occur between adjacent vortex sectors. 
At a sufficiently low temperature
and an optimal screening of Coulomb repulsion by the 
interplane environment (doping) this would lead to a typical equilibrium configuration as depicted in 
Fig.\,\ref{Fig-7} where the intersection points (electronic charge) do not
move relative to one another. A local demolition of this highly ordered 
state would cost a finite amount of energy 
manifested in terms of the (gigantic) 
gaps measured experimentally in the cuprate systems. Applying an electric
field vector with a component parallel to the plane would set into resistivity-free
motion the thus locked electrons. For a macroscopic analogue imagine a stiff table cloth being 
pulled over the table's surface in a friction-free fashion. 
The occurrence of the pseudogap phase would possibly be explained by local
defects in the fabric of Fig.\,\ref{Fig-7} due to insufficient Coulomb
screening and/or too large of a thermal noise (macroscopic vorticity, 
liquid of pointlike defects in 2D).

\section*{Acknowledgments}
We would like to thank Markus Schwarz for useful conversations.


\begin{thebibliography}{10}

\bibitem{Hofmann2005}
R. Hofmann, Int. J. Mod. Phys. A{\bf 20}, 4123 (2005);
Erratum-ibid.A{\bf 21}, 6515 (2006).

\bibitem{GHS2006}
F. Giacosa, R. Hofmann, M. Schwarz, Mod. Phys. Lett. A{\bf 21}, 2709
(2006).\\ 
R. Hofmann, Mod. Phys. Lett. A{\bf 22} 2657 (2007). 

\bibitem{Hofmann2007}
R. Hofmann, arXiv:0710.0962 [hep-th].

\bibitem{NO}
H. B. Nielsen and P. Olesen, Nucl. Phys. B{\bf 61}, 45 (1973).

\bibitem{tHooft1983}
G. 't Hooft, Nucl. Phys. B{\bf 138}, 1 (1978). 

\bibitem{Smith}
S. Smith, M. E. Broucke, and B. A. Francis, in proc. 44th IEEE
Conference on Decision and Control, and European Control Conference
2005, Seville, Spain, Dec. 12-15 (2005). 

\bibitem{GageHamilton}
M. Gage and R. S. Hamilton, J. Differential Geometry {\bf 23}, 69
(1986). 

\bibitem{Grayson}
M. A. Grayson, J. Differential Geometry {\bf 26}, 285
(1987). 

\bibitem{PerelmanI}
G. Perelman, math/0211159.

\bibitem{PerelmanII}
G. Perelman, math/0303109.

\bibitem{PerelmanIII} 
G. Perelman, math/0307245. 

\bibitem{Thurston}
W. P. Thurston, Bull. Amer. Math. Soc. (N.S.) {\bf 6}, no. 3, 357
(1982). 

\bibitem{KKG}
H. V. Klapdor-Kleingrothaus and I. V. Krivosheina,
Mod. Phys. Lett. A{\bf 21}, 1547 (2006).\\ 
H. V. Klapdor-Kleingrothaus, Int. J. Mod. Phys. D{\bf 13}, 2107
(2004).\\ 
H. V. Klapdor-Kleingrothaus, I.V. Krivosheina, A. Dietz, and 
O. Chkvorets, Phys. Lett. B{\bf 586} 198 (2004).

\bibitem{BosiCavalleri2002}
L. Bosi and G. Cavalleri, Nuovo Cim. (note brevi) {\bf 117} B, 243 (2002).  

\bibitem{Zpinch}
F. Giacosa, R. Hofmann, and M. Schwarz, Mod. Phys. Lett. A{\bf 21}, 2709
(2006).

\bibitem{GiacosaHofmann2005}
F. Giacosa and R. Hofmann, Eur. Phys. J. C{\bf 50}, 635 (2007)
[hep-th/0512184].

\bibitem{GraysonII}
M. A. Grayson, Invent. math. {\bf 96}, 177 (1989).

\bibitem{Muller1986} 
J. G. Bednorz and K. A. M\"uller, Z. Phys. B{\bf 64}, 189 (1986). 

\bibitem{Anderson2005}
P. W. Anderson, arXiv:cond-mat/0510053v2.\\ 
P. W. Anderson, Physica C{\bf 460-462}, 3 (2007). 


\end{thebibliography}
\end{document}